# Universal Darwinism as a process of Bayesian inference

John O. Campbell

**Abstract**

Many of the mathematical frameworks describing natural selection are equivalent to Bayes' Theorem, also known as Bayesian updating. By definition, a process of Bayesian Inference is one which involves a Bayesian update, so we may conclude that these frameworks describe natural selection as a process of Bayesian inference. Thus natural selection serves as a counter example to a widely-held interpretation that restricts Bayesian Inference to human mental processes (including the endeavors of statisticians). As Bayesian inference can always be cast in terms of (variational) free energy minimization, natural selection can be viewed as comprising two components: a generative model of an 'experiment' in the external world environment, and the results of that 'experiment' or the 'surprise' entailed by predicted and actual outcomes of the 'experiment'. Minimization of free energy implies that the implicit measure of 'surprise' experienced serves to update the generative model in a Bayesian manner. This description closely accords with the mechanisms of generalized Darwinian process proposed both by Dawkins, in terms of replicators and vehicles, and Campbell, in terms of inferential systems. Bayesian inference is an algorithm for the accumulation of evidence-based knowledge. This algorithm is now seen to operate over a wide range of evolutionary processes, including natural selection, the evolution of mental models and cultural evolutionary processes, notably including science itself. The variational principle of free energy minimization may thus serve as a unifying mathematical framework for universal Darwinism, the study of evolutionary processes operating throughout nature.

**Introduction**

Although Darwin must be counted amongst history's greatest scientific geniuses, he had very little talent for mathematics. His theory of natural selection was presented in remarkable detail, with many compelling examples but without a formal or mathematical framework [1]. Darwin did not think in mathematical terms; he found mathematics repugnant and it comprised only a small part of his Cambridge education [2].

Generally, mathematics is an aid to scientific theories because a theory whose basics are described through mathematical relationships can be expanded into a larger network of predictive implications and the entirety of the expanded theory subjected to the test of evidence. As a bonus, any interpretation of the theory must also conform to this larger network of implications to ensure some consistency.

Natural selection describes the change in frequency or probability of biological traits over succeeding generations. One might suppose that a mathematical description – complete with an insightful interpretation – would be straightforward, but even today this remains elusive. The current impasse involves conceptual difficulties arising from one of mathematics' bitterest interpretational controversies.

That controversy is between the Bayesian and Frequentist interpretations of probability theory. Frequentists assume probability or frequency to be a natural propensity of nature. For instance, the fact that each face of a dice will land with 1/6 probability is understood by frequentists to be a physical property of the dice. On the other hand, Bayesians understand that humans assign probabilities to hypotheses on the basis of the knowledge they have (and the hypotheses they can entertain); thus the probability of each side of a dice is 1/6 because the observer has no knowledge that would favor one face over the other; the only way that no face is favored is for each hypothesis to be assigned the same probability. Furthermore, the value 1/6 is conditioned upon the assumption that there are only six possible outcomes. This means that probabilities are an attribute of a hypothesis or model space – not of the world that is modeled.

The Bayesian framework is arguably more comprehensive and has been developed into the mathematics of Bayesian inference, at the heart of which is Bayes' theorem, which describes how probabilistic models gain knowledge and learn from evidence. In my opinion, the major drawback of the Bayesian approach is an anthropomorphic reliance on human agency, the assumption that inference is an algorithm performed only by humans that possess (probabilistic) beliefs. Despite this interpretational dispute there has been some progress in uniting Bayesian and frequentist mathematics [3].

Despite the lack of mathematics in Darwin's initial formulation it was not long before researchers began developing a mathematical framework describing natural selection. It is an historical curiosity that most of these frameworks involved Bayesian mathematics, yet no interpretations were offered, proposing natural selection as a process of Bayesian inference.

The first step in developing this mathematics was taken during Darwin's lifetime by his cousin, Francis Galton. Galton developed numerous probabilistic techniques for describing the variance in natural traits – as well as for natural selection in general. His conception of natural selection was intriguingly Bayesian; although he may never have heard of Bayes' theorem. Evidence of his Bayesian bent is provided by a visual aid that he built for a lecture on heredity and natural selection given to the Royal Society [4].

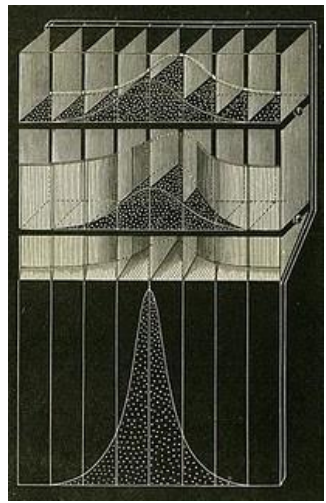

*Figure 1: A device constructed by Francis Galton as an aide in an 1877 talk he gave to the Royal Society. It is meant to illustrate generational change in the distribution of a population's characteristics due to natural selection.*

He used this device (see figure 1 below) to explain natural selection in probabilistic terms. It contains three compartments: a top compartment representing the frequency of traits in the parent population, a middle one representing the application of 'relative fitness' to the child generation and a third representing the normalization of the resulting distribution in the child generation. Beads are loaded in the top compartment to represent the distribution in the parent generation and then are allowed to fall into the second compartment. The trick is in the second compartment, which contains a vertical division, in the shape of the relative fitness distribution. Some of the beads fall behind this division and are 'wasted'; they do not survive and are removed from sight. The remaining beads represent the distribution of the 'survivors' in the child generation.

Galton's device has recently been rediscovered and employed by Stephan Stigler and others in the statistics community as a visual aid, not for natural selection, but for Bayes' theorem. The top compartment represents the prior distribution, the middle one represents the application of the likelihood to the prior, and the third represents the normalization of the resulting distribution. The change between the initial distribution and the final one is the Bayesian update.

R.A. Fisher further developed the mathematics describing natural selection during the 1920s and 1930s. He applied statistical methods to the analysis of natural selection via Mendelian genetics and arrived at the fundamental theorem of natural selection which states [5]:

> *The rate of increase in fitness of any organism at any time is equal to its genetic variance in fitness at that time.*

Although Fisher was a fierce critic of the Bayesian interpretation (which he considered subjective) he pioneered – and made many advances with – the frequentist interpretation.

The next major development in the mathematics of natural selection came in 1970 with the publication of the Price equation, which built on the fundamental theorem of natural selection [6, 7]. Although the Price equation fully describes evolutionary change, its meaning has only recently begun to be unraveled, notably by Steven A. Frank in a series of papers spanning the last couple of decades. Frank's insights into the meaning of the Price equation culminated in a 2012 paper [8] which derives a description of natural selection using the mathematics of information theory.

In my opinion, this paper represents a significant advance in the understanding of evolutionary change as it shifts the interpretation from the objective statistical description of frequentist probability to an interpretation in terms of Bayesian inference. Unfortunately, Frank does not share my appreciation of his accomplishment. While he understands that his mathematics are very close to those of Bayesian inference he does not endorse a Bayesian interpretation but prefers an interpretation in terms of information theory.

**Information and Bayesian inference**

However, the mathematics of information theory and Bayesian probability are joined at the hip, as their basic definitions are in terms of one another. Information theory begins with a definition of information in terms of probability:

$$I(h_i|\text{m}) = -log(P(h_i|\text{m}))$$

Here, we may view $h_i$ as the $i^{th}$ hypothesis or event in a mutually exclusive and exhaustive family of n competing hypotheses comprising a model m. $I(h_i|\text{m})$ is the information gained, under the model, on learning that hypothesis $h_i$ is true. $P(h_i|\text{m})$ is the probability that had previously been assigned by the model that the hypothesis $h_i$ is true. Thus information is 'surprise'; the less likely a model initially considers a hypothesis that turns out to be the case, the more surprise it experiences, and thus the more information it receives.

Information theory, starting with the very definition of information, is aligned with the Bayesian interpretation of probability; information is 'surprise' or the gap between an existing state of knowledge and a new state of knowledge gained through receiving new information or evidence.

The model itself, composed of the distribution of the p($h_i$), may also be said to have an expectation. The information which the model 'expects'' is the weighted average of the information expected by the n individual p($h_i$), which is called the model's entropy.

$$S(H|m) = \sum_{1}^{n} p(h_i|\text{m}) \left(-\log(p(h_i|\text{m}))\right)$$

Entropy is the amount of information that separates a model's current state of knowledge from certainty.

Bayes' theorem follows directly from the axioms of probability theory and may be understood as the implication that new evidence or information holds for the model described by the distribution of the p($h_i$). This theorem states that on the reception of new information (I) by the model (m) the probability of each component hypothesis ($h_i$) making up the model updates according to:

$$P(h_i|I, m) = P(h_i|m) \frac{P(I|h_i m)}{P(I|m)}$$

Bayesian inference is commonly understood as any process which employs Bayes' theorem to accumulate evidence based knowledge [9]: the quantity $P(I|m)$ is called (Bayesian) model evidence and corresponds to the probability of observing some new information, under a particular model, averaged over all hypotheses. This is a crucial quantity in practice and can be used to adjudicate between good and bad models in statistical analysis. It is also the quantity approximated by (variational) free energy – as we will see below. Effectively, this equation provides the formal basis for Bayesian belief updating: in which prior beliefs about the hypotheses $P(h_i|m)$ are transformed into posterior beliefs $P(h_i|I, m)$, which are informed by new information. This updating rests upon the likelihood model; namely the likelihood of observing new information given the *i*-th hypothesis $P(I|h_i m)$. This formalism highlights the information theoretic nature of Bayesian updating – and the key role of a (likelihood) model in accumulating evidence.

We may conclude from this short overview of the relationship between information and Bayesian inference that information has little meaning outside a Bayesian context. Information depends upon a model that assigns probabilities to outcomes and which is updated on the reception of new information. In short, there is no information unless there is something that can be informed. This something is a model.

Thus we see that, contrary to Frank's view, Bayesian inference and information theory have the same logical structure. However, it is instructive to follow Frank's development of the mathematics of evolutionary change in terms of information theory, while keeping in mind his denial of its relationship to Bayesian inference. Frank begins his unpacking of the Price equation by describing the 'simple model' he will develop:

> *A simple model starts with n different types of individuals. The frequency of each type is $q_i$. Each type has $w_i$ offspring, where w expresses fitness. In the simplest case, each type* *is a clone producing $w_i$ copies of itself in each round of reproduction. The frequency of each type after selection is*

$$q'_i = q_i \frac{w_i}{w} \qquad (1)$$

*Where $w = \sum_1^n q_i w_i$ is the average fitness of the trait in the population. The summation is over all of the n different types indexed by the i subscripts.*

Equation (1) is clearly an instance of Bayes' theorem, where the new evidence or information is given in terms of relative fitness and thus Frank's development of this simple model is in terms of Bayesian inference.

While Frank acknowledges an isomorphism between Bayes' theorem and his simple model, he does not find this useful and prefers to describe the relationship as an analogy. He makes the somewhat dismissive remark:

*I am sure this Bayesian analogy has been noted many times. But it has never developed into a coherent framework that has contributed significantly to understanding selection.*

On the contrary, I would suggest that Frank's paper itself develops a coherent framework for natural selection in terms of Bayesian inference. In particular, he highlights the formal relationships between the Price equation (or replicator equation) and Bayesian belief updating (e.g. Kalman Filtering). This is potentially interesting because many results in evolutionary theory can now be mapped to standard results in statistics, machine learning and control theory. Although we will not go into technical details, a nice example here is that Fisher's fundamental theorem corresponds to the increase in Kalman gain induced by random fluctuations (this variational principle is well-known in control theory and volatility theory in economics). Despite this, Frank dismisses Bayesian formulations because they do not appear to bring much to the table. This is understandable in the sense that the mathematics traditionally used to describe natural selection already has a Bayesian form and merely acknowledging this fact does not lead to a new formalism. However, this conclusion might change dramatically if biological evolution was itself a special case of a Universal Darwinism that was inherently Bayesian in its nature. In what follows, we pursue this line of argument by appealing to the variational principle of least free energy.

**Free energy minimization principle**

Baez and Pollard have recently demonstrated the similarities of a number of information-theoretic formulations, including the Bayesian replicator equation, evolutionary game theory, Markov processes and chemical reaction networks, that are applicable to biological systems as they approach equilibrium [10]. In general, any process of Bayesian inference may be cast in terms of (variational) free energy minimization [11, 12] and – in this form – some important interpretative issues gain clarity. This approach has been used by Hinton, Friston and others to describe the evolution of mental states as well as to describe pattern formation and general evolutionary processes. In its most general form, the free energy principle suggests that any weakly-mixing ergodic random dynamical system must be describable in terms of Bayesian inference. This means that the equivalence between classical formulations of evolution and Bayesian updating are both emergent properties of any random dynamical system that sustains measurable characteristics over time (i.e. is ergodic) [13]. This is quite important because it means that evolution is itself an emergent property of any such systems. Although conceptually intriguing, there may be other advantages to treating evolution in terms of minimizing variational free energy. In what follows, I will try to demonstrate this may be true.

In 1970 Ashby and Conant [14] proved a theorem that any regulating mechanism for a complex system that is both successful and simple must be isomorphic with the system being regulated. In other words, it must contain a model of the system being regulated. As no model can be exactly isomorphic to its subject without being a clone and therefore exactly as complex as its subject, this theorem suggests a variational approach may be useful, one which optimizes the difference between the accuracy and the complexity of the model.

This is exactly a form in which the free energy minimization principle may be cast [15]:

$$F(s, u) = D_{KL}[q(\psi|\mu)||p(\psi|m)] - E_q[\log p(s|\psi, m)]$$

Free Energy  =  Complexity  -  Accuracy

Where $\psi$ are hidden states of the world or environment, s are their sensory consequences or samples (that can depend upon action), $\mu$ are internal states and $m$ is the generative model. The distribution q is the current predictions of the states of the environment, the distribution p is the true states of the environment and the KL divergence is a measure of the distance between them. Crucially, free energy can also be expressed in terms of the surprise of sampled consequences:

$$F(s, u) = D_{KL}[q(\psi|\mu)||p(\psi|s,m)] - \log p(s|m)$$

Free Energy    =    relative entropy    +    surprise

This formulation of evolutionary change may appear quite different from that of Bayesian inference as it has a focus on model quality rather than fitness. However, a sustained decrease in free energy (or increase in log model evidence) is equivalent to a decrease in model entropy and therefore contravenes the spirit, if not the letter, of the second law. The letter of that law allows a decrease in entropy for dynamic systems only if an environmental swap is conducted where low entropy inputs are exchanged for high entropy outputs. In short, the second law forbids the existence or survival of low entropy dynamic systems lacking such an ability – an ability that mandates a model of the environment and Bayesian inference under that model. This provides a focus for the model's knowledge accumulation; it must entail knowledge of its environment as well as a strategy to perform the required entropy swaps within that environment. Thus the drive to fitness, which is explicit in the Bayesian formulation, is also implicit in the free energy formulation.

As descriptions of evolutionary processes in terms of free energy minimization have great general applicability it may be useful to consider some specific examples. In biological evolution we can associate the model ($m$) with a genotype. This means the genotype corresponds to the sufficient statistics of the prior beliefs a phenotype is equipped with on entering the world. Keeping in mind that organisms may sense their environments through both chemical and neural means, we may associate sensory exchanges with the environment (s) with adaptive states. Finally, the sufficient statistics of the posterior ($m\mu$) can be associated with a phenotype. In other words, the phenotype embodies probabilistic beliefs about states of its external milieu. This formulation tells us several fundamental things:

i) everything that can change will change to minimize free energy. Here, the only things that can change are the sufficient statistics; namely, the genotype and phenotype. This means there are two optimizations in play: adaptive changes in the phenotype over somatic time (i.e. changes in $m\mu$) and adaptive changes in the genotype over evolutionary time (i.e. changes in $m$).

ii) somatic changes will be subject to two forces: first, a maximization of accuracy that simply maximizes the probability of occupying adaptive states, and second, a minimization of complexity. This minimization corresponds to reducing the divergence between the beliefs about, or model of, (hidden) environmental states ($\psi$) implicit in the phenotype and the prior beliefs implicit in the genotype. In other words, a good genotype will enable the minimization of free energy by equipping the phenotype with prior beliefs that are sufficient to maintain accuracy or a higher probability of adaptive states. Thus the phenotype may be thought of as a type of experiment, which gathers evidence to test prior beliefs; i.e., gathers evidence for its own existence.

iii) changes in the genotype correspond to Bayesian model selection (c.f., natural selection). This simply means selecting models or genotypes that have a low free energy or high Bayesian model evidence. Because the Bayesian model evidence is the probability of an adaptive state given a model or genotype ($p(s|m)$), natural selection's negative variational free energy becomes (free) fitness. At this level of free energy minimization, evolution is in the game of orchestrating multiple (phenotypic) experiments to optimize models of the (local) environment.

Another specific example of the general ability of the free energy minimization principle to describe evolutionary change is in neuroscience where it is fairly easy to demonstrate the centrality of this principle in explaining evolutionary, developmental and perceptual processes in a wide range of mental functions [11]. The brain produces mental models which combine sensory information concerning the state of the environment, with possible actions with which the organism may intervene. The initiation of an action is a kind of experiment in the outside world testing the current beliefs about its hidden states. The overall drive of the free energy principle is to reduce the model complexity, while maximizing its accuracy in achieving the predicted outcome. Crucially, the ensuing self-organization can be seen at multiple levels of organization; from dendritic processes that form part of the single neuron – to entire brains. The principles are exactly the same, the only thing that changes is the way that the model is encoded (e.g., with intracellular concentrations of various substrates – or neuronal activity and connectivity in distributed brain systems). This sort of formulation has also been applied to self-organization and pattern formation when multiple systems jointly minimize their free energy (for example, in multi-agent games and morphogenesis at the cellular level).

Clearly, the application of variational (Bayesian) principles to ecological and cellular systems means we have to abandon the notion that only humans can make inferences. We will take up this theme below and see how freeing oneself from the tyranny of anthropomorphism leads us back to a universal Darwinism.

The free energy minimization principle may also be applied to processes of cultural evolution. A compelling example here is the evolution of scientific understanding itself. Science develops hypotheses or theoretical models of natural phenomena. These models are used to design experiments in the real world and the results of the experiment are used to update the probability of each hypothesis composing the model according to Bayes' theorem. In the process free energy is minimized through a balance which reduces the model's complexity (Occam's razor) while increasing the model's predictive accuracy and explanatory scope.

The evolutionary interaction between models and the systems they model, as described by the free energy minimization principle, may be applicable to additional natural phenomena beyond the examples above. Several attempts have been made to describe universal Darwinism in such terms. We have previously noted the wide range of scientific subject matter that has been identified within the literature as Darwinian processes – and have offered an interpretation in terms of inferential systems [16]; an interpretation closely related to that of the free energy minimization principle. Richard Dawkins offered a description of biological evolution in terms of replicators and vehicles [17], a description which Blackmore and Dennett have generalized to interpret universal Darwinism [18, 19]. That description may also be understood as an interplay between internal models (replicators) and the experience of the 'experiments' (vehicles) which they model in the external world.

The Price equation describing evolutionary change may be cast in a form which distinguishes between change due to selection and transmission. Changes due to selection tend to decrease model variation whereas changes due to transmission or copying of the model serve to increase variation. The transmission changes of biological models are often in the form of genetic mutations [20]. From the perspective of universal Darwinism, we might expect a mechanism capable of increasing model variation within non-biological evolutionary processes that is analogous to biological mutation. As an example we might consider the process of evolutionary change in scientific models during transmission. These may appear less clear; there is less consensus on how new and sometimes improved scientific models are generated. It may seem this process has little in common with the somewhat random and undirected process of biological mutation.

The mental process by which researchers arrives at innovative models is largely hidden and might be considered closer to an art form than algorithmic but the development of inferential/Darwinian evolutionary computational processes have demonstrated a strong ability to discover innovative models in agreement with the evidence [21, 22]. In some instances, these evolutionary approaches have inferred successful models for systems which have long eluded researchers [23].

**The arena of Bayesian inference**

The reluctance of many researchers to endorse a Bayesian interpretation of evolutionary change may be somewhat puzzling. One reason for this is a peculiarity, and I would suggest a flaw, in the usual Bayesian interpretation of inference that renders it unfit as a description of generalized evolutionary change. The consensus Bayesian position is that probability theory only describes inferences made by humans. As E.T. Jaynes put it [24]:

> *it is...the job of probability theory to describe human inferences at the level of epistemology.*

Epistemology is the branch of philosophy that studies the nature and scope of knowledge. Since Plato the accepted definition of knowledge within epistemology has been 'justified true beliefs' held by humans. In the Bayesian interpretation 'justified' means justified by the evidence. 'True belief' is the degree of belief in a given hypothesis which is justified by the evidence; it is the probability that the hypothesis is true within the terms of the model. Thus knowledge is the probability, based on the evidence, that a given belief or model is true. I have proposed a technical definition of knowledge as $2^{-S}$ where S is the entropy of the model [16].

A perhaps interesting interpretation of this definition is that knowledge occurs within the confines of entropy or ignorance. For example, in a model composed of a family of 64 competing hypotheses, where no evidence is available to decide amongst them, we would assign a probability of 1/64 to each hypothesis. The model has an entropy of 6 bits and has knowledge of $2^{-6} = 1/64$. Let's say some evidence becomes available and the model's entropy or ignorance is reduced to 3 bits. Then the knowledge of the updated model is 1/8, equivalent to the entropy of a model composed of only 8 competing hypotheses that is maximally ignorant, which has no available evidence. The effect which evidence has on the model is to increase its knowledge by reducing the scope of its ignorance.

It is unfortunate that both Bayesian and Frequentist interpretations deny the existence of knowledge outside of the human realm because it forbids the application of Bayesian inference to phenomena other than models conceived by humans, it denies that knowledge may be accumulated in natural processes unconnected to human agency and it acts as a barrier in realizing our close relationship to the rest of nature. Thus even though natural selection is clearly described in terms of the mathematics of Bayesian inference, neither Bayesians such as Jaynes nor frequentists such as Frank can acknowledge this fact due to another hard fact: natural selection is not dependent upon human agency. In both their views this may rule out a Bayesian interpretation.

I believe that the correct way out of this conundrum is to simply acknowledge that in many cases inference is performed by non-human agents as in the case of natural selection and that inference is an algorithm which we share with much of nature. The genome may for instance be understood as an example of a non-human conceived model involving families of competing hypotheses in the form of competing alleles within the population. Such models are capable of accumulating evidence-based knowledge in a Bayesian manner. The evidence involved is simply the proportion of traits in ancestral generations which make it into succeeding generations. *In other words, we just need to broaden Jaynes' definition of probability to include non-human agency in order to view natural selection in terms of Bayesian inference.*

In this view the accumulation of knowledge is a preoccupation we share with the rest of nature. It allows us to view nature as possessing some characteristics, such as surprise and expectations, previously thought by many as unique to humans or at least to animals. For instance, all organisms 'expect' to find themselves in the type of environment for which they have been adapted and are 'surprised' if they don't.

**Universal Darwinism**

Bayesian probability, epistemology and science in general tend to draw a false distinction between the human and non-human realms of nature. In this view the human realm is replete with knowledge and thus infused with meaning, purpose and goals, and Bayesian inference may be used to describe its knowledge-accumulating attributes. On the other hand, the non-human realm is viewed as devoid of these attributes and thus Bayesian inference is considered inapplicable.

However, if we recognize expanded instances, such as natural selection, in which nature accumulates knowledge then we may also recognize that Bayesian inference, as well as equivalent mathematical forms, provides a suitable mathematical description in both realms. Evolutionary processes, as described by the mathematics of Bayesian inference, are those which accumulate knowledge for a specific purpose, knowledge required for increased fitness, for increased chances of continued existence. Thus the mathematics implies purpose, meaning and goals, and provides legitimacy for Daniel Dennett's interpretation of natural selection in those terms [19]. If we allow an expanded scope for Bayesian inference, we may view Dennett's poetic interpretation of Darwinian processes as having support from its most powerful mathematical formulations.

An important aspect of these mathematics is that they apply not only to natural selection but also to any generalized evolutionary processes where inherited traits change in frequencies between generations. As noted in a cosmological context by Conlon and Gardner [25]:

> *Specifically, Price's equation of evolutionary genetics has generalized the concept of selection acting upon any substrate and, in principle, can be used to formalize the selection of universes as readily as the selection of biological organisms.*

At the core of Bayesian inference, underlying both the Price equation and the principle of free energy minimization we find an extremely simple mathematical expression: Bayes' theorem:

$$q'_i = q_i \frac{w_i}{w}$$

Simply put this equality says that the probabilities assigned to the hypotheses of a probabilistic model are updated by new data or experience according to a ratio, that of the probability of having the experience given that the specific hypothesis is correct to the average probability assigned by the model to having that experience. Those hypotheses supported by the data, those that assign greater than average probability to having the actual experience, will be updated to greater values and those hypotheses not supported by the data will be updated to lesser values. This simple equation describes the accumulation of evidence-based knowledge concerning fitness.

When Bayes' theorem is used to describe an evolutionary process the ratio involved is one of relative fitness, the ratio of the fitness of a specific form of a trait to the average fitness of all forms of that trait. It is thus extremely general in describing any entity able to increase its chances of survival or to increase its adaptiveness. When cast in terms of the principle of free energy minimization some further implications of this simple equation are revealed (see above).

In a biological evolutionary context, the Price equation is traditionally understood as the mathematics of evolutionary change. However, the Price equation may be derived from a form of Bayes' theorem [26, 8, 27] which means it describes a process of Bayesian inference, a very general form of Bayesian inference which according to Gardner [26] applies to any group of entities that undergo transformations in terms of a change in probabilities between generations or iterations. Even with this great generality it provides a useful model as it partitions evolutionary change in terms of selection and transmission [7].

There are numerous examples of these equivalent mathematical forms used in the literature to describe evolutionary change across a wide scope of scientific subject matter, specifically evolutionary change in biology [8, 26], neuroscience [11, 28] and culture [29, 30, 31, 32].

It is interesting to speculate on the similarity of these mathematical forms to those which may be used to describe quantum physics. Quantum physics is also based upon probabilistic models which are updated by information received through interactions with other entities in the world.

Wojciech Zurek, the founder of the theory of quantum Darwinism [33], notes that the update of quantum states may be understood in terms of ratios acting to update probabilistic models [34].

> *Using this connection, we then infer probabilities of possible outcomes of measurements on S from the analogue of the Laplacian 'ratio of favorable events to the total number of equiprobable events', which we shall see in Section V is a good definition of quantum probabilities for events associated with effectively classical records kept in pointer states.*

Unfortunately, many who have attempted to interpret quantum theory in terms of Bayesian inference, such as Caves, Fuchs and Schack [35], have endorsed a common anthropomorphic Bayesian flaw and conclude that the probabilities involved with quantum phenomena are a 'personal judgment' [36], and thus that the inferences involved take place within a human brain. A conceptual shift acknowledging that inference is a natural algorithm which may be performed in processes outside of the human brain may go some way to allowing quantum Darwinism to be understood as a process of Bayesian inference conducted at the quantum level [37].

A vast array of phenomena is subject to evolutionary change and describable by the equivalent mathematical forms discussed here. These forms interpret evolutionary change as based on the accumulation of evidence-based knowledge. Conversely, many instances of evidence-based knowledge found in nature are describable using this mathematics. We might speculate that all forms of knowledge accumulation found in nature may eventually find accommodation within this paradigm. Certainly, the theorem proved by Cox [38] identifies Bayesian inference as the unique method by which models may be updated with evidence.

It is somewhat ironic that in 1935 R.A. Fisher wrote [39]:

> *Inductive inference is the only process known to us by which essentially new knowledge comes into the world.*

Of course he was referring to experimental design and considered it unnecessary to specify that he was referring only to human knowledge. Probably he assumed that no other repositories of knowledge exist. The stage may now be set for us to understand his assertion as literally true in its full generality.

Ultimately the scope and interpretation of universal Darwinism, the study of phenomena which undergoes evolutionary change, will depend on the mathematical model underlying it. Those phenomena which are accurately and economically described by the mathematics must be judged to be within the scope of universal Darwinism. Given the great generality and substrate independence of current mathematical models, a unification of a wide range of scientific subject matters within this single paradigm may be possible.

**Acknowledgements**